\documentclass[showpacs,prl,onecolumn,aps,superscriptaddress,preprintnumbers,nofootinbib]{revtex4}
\usepackage[T1]{fontenc}
\usepackage[latin9]{inputenc}
\usepackage{graphicx,hyperref}

\usepackage{amsmath,amssymb}
\usepackage{epsfig}
\usepackage{graphicx}
\usepackage{amsmath}
\usepackage{amsfonts}
\def\slashchar#1{\setbox0=\hbox{$#1$}     		
   \dimen0=\wd0                                 	
   \setbox1=\hbox{/} \dimen1=\wd1               	
   \ifdim\dimen0>\dimen1                        	
      \rlap{\hbox to \dimen0{\hfil/\hfil}}      	
      #1                                        	
   \else                                        	
      \rlap{\hbox to \dimen1{\hfil$#1$\hfil}}   	
      /                                         	
   \fi}

\renewcommand{\vec}{\boldsymbol}
\newcommand{\beq}{\begin{equation}}
\newcommand{\eeq}{\end{equation}}
\newcommand{\bea}{\begin{eqnarray}}
\newcommand{\eea}{\end{eqnarray}}
\newcommand{\baa}{\begin{array}}
\newcommand{\eaa}{\end{array}}

\def\eq#1{{Eq.~(\ref{#1})}}

\def\fig#1{{Fig.~\ref{#1}}}

\newcommand{\bas}{\bar{\alpha}_S}

\newcommand{\nn}{\nonumber}

\newcommand{\h}{\frac{1}{2}}

\newcommand{\x}{\vec{x}}

\newcommand{\Lb}{\left(}
\newcommand{\Rb}{\right)}

\renewcommand{\vec}[1]{\boldsymbol{#1}}

\begin{document}
\title{ Non-linear evolution in the re-summed next-to-leading order of perturbative QCD: \\confronting the experimental data}
\author{Carlos Contreras}
\email{carlos.contreras@usm.cl}
\affiliation{Departamento de F\'isica, Universidad T\'ecnica Federico Santa Mar\'ia,  Avda. Espa\~na 1680, Casilla 110-V, Valpara\'iso, Chile}
\author{ Eugene ~ Levin}
\email{leving@tauex.tau.ac.il, eugeny.levin@usm.cl}
\affiliation{Departamento de F\'isica, Universidad T\'ecnica Federico Santa Mar\'ia,  Avda. Espa\~na 1680, Casilla 110-V, Valpara\'iso, Chile}
\affiliation{Centro Cient\'ifico-
Tecnol\'ogico de Valpara\'iso, Avda. Espa\~na 1680, Casilla 110-V, Valpara\'iso, Chile}
\affiliation{Department of Particle Physics, School of Physics and Astronomy,
Raymond and Beverly Sackler
 Faculty of Exact Science, Tel Aviv University, Tel Aviv, 69978, Israel}

\author{Michael Sanhueza}
\email{michael.sanhueza.roa@gmail.com}
\affiliation{Departamento de F\'isica, Universidad T\'ecnica Federico Santa Mar\'ia,   Avda. Espa\~na 1680, Casilla 110-V, Valpara\'iso, Chile}
\date{\today}

\keywords{BFKL Pomeron,  CGC/saturation approach, solution to non-linear
 equation, deep inelastic/Users/leving/Library/Containers/com.apple.mail/Data/Library/Mail Downloads/4CEE7639-2DA9-413A-81BD-72B039E4A25E/Screenshot[1].png
 structure function}
\pacs{ 12.38.Cy, 12.38g,24.85.+p,25.30.Hm}
\begin{abstract}
In this paper  we  compare  the experimental HERA  data with  the next-to-leading order approach (NLO) of Ref.~\cite{CLMS}. This approach  includes the re-summed  NLO corrections to the kernel of the evolution equation, the correct asymptotic behaviour in the NLO at $\tau = r^2 Q^2_s \,\gg\,1$; the impact parameter dependence of the saturation scale in accord with the Froissarrt theorem as well as the non-linear corrections. In this paper, we  successfully describe the experimental data with the quality,  which is not worse,  than in the leading order fits with larger number of the phenomenological parameters. It is demonstrated, that the data could be described,  
 taking into account both the diffusion on $\ln(k_T)$, which stems from perturbative QCD, and the Gribov's diffusion in  impact parameters.   It is shown   an ability to describe the data at rather large values of $\bas$
.
  \end{abstract}
\maketitle

\vspace{-0.5cm}
\tableofcontents






\section{ Introduction}

The goal of this paper is to compare with the experimental (HERA) data  the next-to-leading order approach (NLO) of Ref.~\cite{CLMS}. In  Ref.~\cite{CLMS},  we develop the approach in which we include 
 the re-summation procedure, suggested in
 Refs.~\cite{SALAM,SALAM1,SALAM2}, to fix the BFKL kernel in the NLO. 
 In particular, we introduce  the rapidity variable, which plays the role of the  ``evolution time",  in the same way as in Ref.~\cite{DIMST}. However, 
we suggest a different way to account for  the non-linear corrections, 
than in Ref.~\cite{DIMST}, which leads to additional change of the NLO kernel of the evolution equation.  The advantage of our kernel of the BFKL equation~\cite{BFKL,LIP},  is that the scattering amplitude satisfies the high energy limits, which follows from the approach of Ref.~\cite{LETU} (see Refs.~\cite{CLMP,XCWZ}) to the NLO Balitsky-Kovchegov (BK) ~\cite{BK}  evolution~\cite{NLOBK0,NLOBK01,NLOBK1,NLOBK2,JIMWLKNLO1,JIMWLKNLO2,JIMWLKNLO3}.

We firmly believe that finding the correct NLO approximation for the non-linear evolution is one of the most important and urgent problem in the theoretical description of the high energy scattering. Indeed,  in the Colour Glass Condensate(CGC) approach, which  is the only candidate for an
 effective theory at high energies (see Ref.~\cite{KOLEB} for a review),   the two essential parameters, that determine the high energy
 scattering, calculated in leading order of perturbative QCD ~\cite{BFKL,GLR,MUQI,MV,JIMWLK1,JIMWLK2,JIMWLK3,JIMWLK4,JIMWLK5,JIMWLK6,BK} turns out to be in  an apparent contradiction with the experimental data. The first one is 
 the BFKL Pomeron~\cite{BFKL} intercept, which is equal
 to $ \,2.8\, \bas$ and leads to the energy behaviour of the scattering
 amplitude $N \propto \exp\left(2.8\,\bas \ln(\frac{1}{x})\right)$.  The second is 
the
 energy behaviour of the new dimensional scale: saturation momentum  $Q^2_s
 \propto \exp\left(4.88\,\bas \ln(\frac{1}{x})\right)$. Both  show the 
increase in the leading order CGC approach, which cannot be reconciled 
with the available experimental data. So, the large NLO corrections appear
  as the only way out,  now as well as  two decades ago.

  In the next section we will outline the main results  of Ref.~\cite{CLMS} and will specify our theoretical description of the dipole scattering amplitude.  However,  the current  stage of our theoretical understanding  of non-perturbative QCD is such,  that we have  to build a model. We need to take into account the non-perturbative corrections that will reproduce the correct, exponentially decreasing at large impact parameters ($b$) scattering amplitude. It has been demonstrated in Refs.~\cite{KW1,KW2,KW3,FIIM}, that the CGC equations ~\cite{JIMWLK1,JIMWLK2,JIMWLK3,JIMWLK4,JIMWLK5,JIMWLK6,BK}  as well as all other approaches, based on perturbative QCD,  lead to the amplitude that increases as a power of energy, resulting in the violation of the Froissart theorem  ~\cite{FROI}\footnote{It should be noted that NLO corrections to the kernel of CGC evolution equations does not change the power-like behaviour of the scattering 
  amplitude but improve the situation moving the violation of the Froissart theorem to larger values of $b$     ~\cite{BEST1,BEST2,CCM,BCCM,CLM}.}. Unfortunately, without a theoretical control on non-perturbative QCD we have to use a phenomenological  approach to model the large $b$ behaviour. In this paper we will exploit two approaches: 
  
  \begin{enumerate}
  \item  the non-perturbative behaviour of the saturation scale, which we will parameterize as follows:
   \beq \label{QSB}
 Q^2_s\Lb b , Y\Rb \,\propto\, \Lb S\Lb b, m \Rb\Rb^{\frac{1}{\bar \gamma}}
 \eeq
     where     $S\Lb b \Rb $ is the Fourier  image of $ S\Lb Q_T\Rb = 1/\Lb 1 +    \frac{Q^2_T}{m^2}\Rb^2$        and the value of $\bar \gamma$ we will discuss below.  In the vicinity of the saturation scale   such $b$  dependance    results in the large  $b$-dependence of the scattering amplitude,  which is   proportional to $\exp\Lb - m\, b\Rb$ at $b \gg 1/m$,  in  accordance of the Froissart theorem ~\cite{FROI}. In addition, we reproduce the large $Q_T$ dependence of this amplitude proportional to $Q^{-4}_T$ which follows from the perturbative QCD calculation ~\cite{BRLE}. Theoretically the fact that we can absorb the non-perturbative $b$-dependence in $Q_s(Y,b)$ (see  \eq{QSB}  for example), follows from the semi-classical approach to BK equation~\cite{BKL} and has been widely used in all, so called, saturation models ~\cite{SATMOD0,SATMOD1,SATMOD2,IIM,SATMOD3,SATMOD4,SATMOD5,SATMOD6,SATMOD7,SATMOD8, SATMOD9,SATMOD10,SATMOD11,SATMOD12,SATMOD13,SATMOD14,SATMOD15,SATMOD16,SATMOD17,CLP,CLMP}.

   \item  in Refs.~\cite{LERY1,LERY2,LETAN,QCD2,KHLE,KKL,BLT,LEPION,KAN,GOLEB}  it is made an attempt to incorporate in the BFKL equation the
 Gribov's   diffusion ~\cite{GRIB}  in  impact parameter ($b$). As the result of this   tthe following formula for the saturation scale was suggested ~\cite{LEPION,GOLEB}:
    \beq \label{QSB1}
 Q^2_s\Lb b , Y\Rb \,\propto\, \exp\Lb - \frac{3}{4}\,{\cal Z}\Rb
 ~~~~\mbox{with}~~~~  
     {\cal Z} \,=\,\Lb \frac{b^4}{4 {\alpha'}^2_{\rm eff}\,Y}\Rb^{1/3}
    \eeq   
where $ { \alpha'}^2_{\rm eff} $ is a new dimensional non-perturbative parameter which controls the Gribov's diffusion.
 
    \end{enumerate}     
  
  The paper is organized as follows. In section II  we will give a brief review of the  approach that we have developed in Ref. ~\cite{CLMS}. Next we will discuss   the procedure of obtaining the NLO kernel of the BFKL equation based on the anomalous dimensions  for re-summed NLO corrections,
  which is suggested in Refs. ~\cite{SALAM,SALAM1,SALAM2,KMRS}.  Then we will consider the modification that we need to introduce in the NLO kernel to account for correct behaviour of the scattering amplitude at high energy and the structure of the  non-linear equation. In section III we specify our approach, which we use for describing the experimental data. In particular, we introduce the phenomenological parameters, which have to be calculated in the non-perturbative QCD approach, and discuss their physical meaning and the range of possible values.  In section IV we collect the results of the fit of the experimental data on DIS. We summarize our results in the conclusion.

\section{Leading twist approximation for  non-linear evolution in the NLO - a recap}
\subsection{ Re-summed anomalous dimensions in NLO and the kernel of linear evolution}
The general solution to the linear equation has the following form:
\beq \label{GENSOL}
N\Lb Y, \vec{r},\vec{b}\Rb\,\,\,=\,\,\int^{\epsilon + i \infty}_{\epsilon - i \infty}\frac{d \gamma}{2\,\pi\,i} e^{\omega\Lb \gamma\Rb\,Y} \phi_\gamma\Lb \vec{r}, \vec{b},\vec{R}\Rb\,\,\phi_{in}\Lb \gamma, R\Rb\,\,\,=\,\, \int^{\epsilon + i \infty}_{\epsilon - i \infty}\frac{d \omega}{2\,\pi\,i} e^{\omega\,Y} \phi_{\gamma(\omega)} \Lb \vec{r}, \vec{b},\vec{R}\Rb\,\,\phi_{in}\Lb \omega, R\Rb
 \eeq 
where $N$ is the scattering amplitude of the dipole with size  $r$  at  the impact parameter $b$. $Y$ is the rapidity of the dipole. $\phi_\gamma\Lb \vec{r}, \vec{b}\Rb$ is the eigenfunction of the BFKL equation which has the general form:
\bea 
\phi_\gamma\Lb \vec{r} , \vec{R}, \vec{b}\Rb\,&=&\,\Lb \frac{ r^2\,R^2}{\Lb \vec{b}  + \h(\vec{r} - \vec{R})\Rb^2\,\Lb \vec{b}  -  \h(\vec{r} - \vec{R})\Rb^2}\Rb^\gamma\,\,=\,\,\,e^{\gamma\,\xi} \label{EIGENF}\\
\mbox{with}~~~ \xi\,&=&\,\ln\Lb \frac{ r^2\,R^2}{\Lb \vec{b}  + \h(\vec{r} - \vec{R})\Rb^2\,
\Lb \vec{b}  -  \h(\vec{r} - \vec{R})\Rb^2}\Rb \label{XI}
\eea
 where $R$ is the size of the target. $\phi_{in}$ can be found from the initial condition at $Y=0$.
 
 The eigenvalues $\omega\Lb \gamma\Rb$ in the NLO has been calculated in
 Refs. ~\cite{BFKLNLO,BFKLNLO1} and have the following form:
\beq \label{KERNLO1}
\omega_{\rm NLO}\Lb \bas,  \gamma\Rb\,\,=\,\,\bas\,\chi^{LO}\Lb \gamma \Rb\,\,+\,\,\bas^2\,\chi^{NLO}\Lb  \gamma\Rb
\eeq
The explicit form of $\chi^{NLO}\Lb  \gamma\Rb$ is given in
 Ref.~\cite{BFKLNLO}. However, \eq{KERNLO1} has singularities at $\gamma \,\to\,1$ which has been re-summed taking into account the  high order corrections   in Refs.
   ~\cite{SALAM,SALAM1,SALAM2,KMRS}.  Finally, $\omega_{\rm NLO}\Lb \bas,  \gamma\Rb$  has the form
         ~\cite{SALAM,SALAM1,SALAM2}:
\beq \label{KERNLOR}
\omega_{\rm NLO}\Lb \bas,  \gamma\Rb\,=\,\bas \Lb \chi_0\Lb\omega_{\rm NLO}, \gamma\Rb\,+\,\omega_{\rm NLO} \,\frac{\chi_1\Lb \omega_{\rm NLO}, \gamma\Rb}{ \chi_0\Lb\omega_{\rm NLO}, \gamma\Rb}\Rb
\eeq
where
\beq \label{CHI0}
\chi_0\Lb\omega, \gamma\Rb\,\,=\,\,\chi^{LO}\Lb \gamma\Rb \,-\,\frac{1}{ 1 \,-\,\gamma}\,+\,\frac{1}{1\,-\,\gamma\,+\,\omega}
\eeq
and
\bea \label{CHI1}
&&\chi_1\Lb\omega, \gamma\Rb\,\,=\\
&&\,\,\chi^{NLO}\Lb \gamma\Rb\,+\,F\Lb \frac{1}{1 - \gamma}\,-\,\frac{1}{1\,-\,\gamma\,+\,\omega}\Rb\,+\,\frac{A_T\Lb \omega\Rb \,-\,A_T\Lb 0 \Rb}{\gamma^2} \,+\, \frac{A_T\Lb \omega\Rb - b}{\Lb 1\,-\,\gamma\,+\,\omega\Rb^2}\,-\,\frac{A_T\Lb 0\Rb - b}{\Lb 1\,-\,\gamma\Rb^2}\nn
 \eea
Functions $\chi^{NLO}\Lb \gamma\Rb$ and $A_T\Lb \omega\Rb$ as well as 
 the constants ($F$ and $b$),  are defined in 
Refs. ~\cite{SALAM,SALAM1,SALAM2}, while $\chi^{LO}\Lb \gamma
 \Rb$  has the following form:
 \beq \label{CHI}
\omega_{\rm LO}\Lb \bas, \gamma\Rb\,\,=\,\,\bas\,\chi^{LO}\Lb \gamma
 \Rb\,\,\,=\,\,\,\bas \Lb 2 \psi\Lb 1\Rb \,-\,\psi\Lb \gamma\Rb\,-\,\psi\Lb
 1 - \gamma\Rb\Rb
\eeq
 However, in Ref. ~\cite{KMRS}  the economic form  of $\chi_1\Lb \omega,\gamma\Rb$ is given,   which coincides
  with \eq{CHI1} to within  $7\%$: 
  \beq \label{KMRSOM}
\omega^{\rm KMRS} \,=\,\bas\Lb 1 - \omega^{\rm KMRS}\Rb \Lb \frac{1}{\gamma}  + \frac{1}{1 - \gamma + \omega^{\rm KMRS}}\,+\,\underbrace{\Lb 2 \psi(1) - \psi\Lb 2 - \gamma\Rb -  \psi\Lb 1 + \gamma\Rb\Rb}_{\mbox{ high twist contributions}}\Rb
\eeq
One  can see that $\gamma(\omega) \to 0$ when $\omega \to 1$ as follows
 from energy conservation.
 
\eq{KERNLOR} for $\gamma \to 1$ has the form:
\beq \label{GANLOS}
\omega\,\,=\,\,\frac{\bas}{1 - \gamma + \omega};
\eeq
which leads to 
\beq \label{OMNLO}
\omega\Lb \gamma\Rb\,\,=\,\,\h\Lb-\Lb 1  - \gamma\Rb\,+\,\sqrt{4\,\bas \,+\,\Lb 1 - \gamma\Rb^2}\Rb
\eeq
As it is shown in Ref. ~\cite{CLMS} \eq{OMNLO} corresponds to the kernel of Ref.  ~\cite{DIMST}.
Resolving   \eq{GANLOS}  with respect to $\gamma$ ~\cite{ASV,CLMS} we obtain: 
  \beq \label{DLANLO1}
 1 \,-\,\gamma\,\,\,=\,\,\,\frac{\bas}{\omega} \,\,-\,\,\omega
 \eeq
 \eq{DLANLO1} gives the simple equation
\beq \label{DLANLO2}
 \frac{\partial}{\partial \,\eta}\tilde{N}\Lb \xi', \eta; b \Rb =
 \bas \int^{\xi'} d \xi'' \,\tilde{N}\Lb \xi'',\eta; b\Rb\,;~~~~~~~~~
 \frac{\partial^2}{\partial \,\eta\,\partial\,\xi'}\tilde{N}\Lb \xi',
 \eta; b \Rb\,\,=\,\, \bas \, \tilde{N}\Lb \xi',\eta; b\Rb\,,\eeq 
 for the amplitude $\tilde{N}\Lb \xi', \eta; b \Rb\,=\, N\Lb \xi', \eta; b \Rb \Big{/}r^2$. $\eta = Y - \xi'$ is a new energy variable, which corresponds correct time ordering in double log approximation (DLA) ~\cite{DIMST}. In \eq{DLANLO1} $\xi' \,=\,- \,\xi$.
 
 \subsection{Non-linear equation and the feedback to the kernel of the linear evolution}
 
 The general structure of the non-linear Balitsky-Kovchegov equation ~\cite{BK} has the following form:
 
 \bea \label{BK}
&&\frac{\partial}{\partial Y}N\Lb \vec{x}_{10}, \vec{b} ,  Y; R \Rb = \nn\\
&&\bas\!\! \int \frac{d^2 \vec{x}_2}{2\,\pi}\,K\Lb \vec{x}_{02}, \vec{x}_{12}; \vec{x}_{10}\Rb \Bigg(N\Lb \vec{x}_{12},\vec{b} - \h \vec{x}_{20}, Y; R\Rb + 
N\Lb \vec{x}_{20},\vec{b} - \h \vec{x}_{12}, Y; R\Rb - N\Lb \vec{x}_{10},\vec{b},Y;R \Rb\nn\\
&&-\,\, N\Lb \vec{x}_{12},\vec{b} - \h \vec{x}_{20}, Y; R\Rb\,N\Lb \vec{x}_{20},\vec{b} - \h \vec{x}_{12}, Y; R\Rb\Bigg)
\eea
where $\vec{x}_{i k}\,\,=\,\,\vec{x}_i \,-\,\vec{x}_k$  and $ \vec{x}_{10}
 \equiv\,\vec{r}$, $\vec{x}_{20}\,\equiv\,\vec{r}' $ and $\vec{x}_{12}
 \,\equiv\,\vec{r}\,-\,\vec{r}'$.  $Y$ is the rapidity of the scattering
 dipole and $\vec{b}$ is the impact factor. $K\Lb \vec{x}_{02}, \vec{x}_{12};
 \vec{x}_{10}\Rb$ is the kernel of the BFKL equation. In our approach we wish to preserve this form, but include the NLO corrections to the kernel. In particular, we would like to include the reggeization term in \eq{BK}, which contribute to the linear equation, but has been neglected in DLA (see \eq{DLANLO2}). It should be stressed that only keeping this term we can provide the correct asymptotic behaviour at large $Y$: $N\,\,\to\,\,1$.
 
 Therefore, the linear equation takes the form:
  \bea \label{DLANLO3}\
&&\frac{\partial}{\partial \eta}N\Lb \vec{x}_{10}, \vec{b} ,  Y; R \Rb = \nn\\
&&\bas\!\! \int \frac{d^2 \vec{x}_2}{2\,\pi}\,K\Lb \vec{x}_{02}, \vec{x}_{12}; \vec{x}_{10}\Rb \Bigg\{N\Lb \vec{x}_{12},\vec{b} - \h \vec{x}_{20}, Y; R\Rb + 
N\Lb \vec{x}_{20},\vec{b} - \h \vec{x}_{12}, Y; R\Rb - N\Lb \vec{x}_{10},\vec{b},Y \Rb\Bigg\}\nn\\   
 && \xrightarrow{ x_{12} \sim x_{02}\,\,\gg\,\,x_{01}}\,\,\h\,\bas\,x^2_{01}\,\int_{x^2_{01}} \frac{d x^2_{12}}{x^4_{12} }  \Bigg( 2\,N\Lb \vec{x}_{12},\vec{b} , Y; R\Rb \,\,-\,\, N\Lb \vec{x}_{01},\vec{b} , Y; R\Rb \Bigg)
 \eea
 and \eq{DLANLO1} can be re-written in the form:
  \beq \label{DLALONL2}
 \frac{\partial^2}{\partial \eta\,\partial\,\xi'}\tilde{N}\Lb \xi',\eta \Rb \,\,=\,\,\bas \,\tilde{N}\Lb \xi',\eta \Rb \,\,-\,\,\h\,\bas
  \frac{\partial}{\partial\,\xi'}\tilde{N}\Lb \xi',\eta \Rb 
  \eeq 
 
  For the dipole amplitude, $N\Lb \xi',Y\Rb$, \eq{DLANLO1} takes
 the following form:

 \beq \label{DLALONL20}
 \frac{\partial^2}{\partial Y\,\partial\,\xi'} N \Lb \xi',Y \Rb \,\,+\,\, \frac{\partial}{\partial Y} N \Lb \xi',Y \Rb\,\,=\,\,\h\,\bas \,N\Lb \xi',Y \Rb \,\,-\,\,\h\,\bas
  \frac{\partial}{\partial\,\xi'} N\Lb \xi',Y \Rb 
  \eeq  
  which leads to the eigenvalue $\omega(\gamma)$:
   \beq \label{DLALONL21}  
  \omega\Lb \gamma\Rb\,\,=\,\,\h\,\bas\,\frac{1 \,+\,\gamma}{1\,-\,\gamma}
  \eeq
  Using \eq{KMRSOM} in the vicinity of $\gamma \to 1$ one can see that \eq{DLALONL21} takes the form~\cite{CLMS}:
  \beq \label{DLALONL22}  
  \omega\Lb \gamma\Rb\,\,=\,\,\h\,\bas\,\frac{1 \,+\,\gamma}{1\,+\,\bas\,-\,\gamma}
  \eeq  
   Using the general equation to determine the critical anomalous dimension and the energy behaviour of the saturation scale (see review ~\cite{KOLEB}):
   \beq \label{OMCR}
\lambda_\eta\,\,=\,\,\frac{\omega\Lb \bar{\gamma}_\eta\Rb}{\bar{\gamma}_\eta}\,\,=
\,\,-\, \frac{d \omega\Lb \bar{\gamma}_\eta \Rb}{d \bar{\gamma}_\eta};~~~\ln\Lb \frac{Q^2_s\Lb Y\Rb}{Q^2_s\Lb Y=0\Rb} \Rb\,\,=\,\,\lambda_\eta \,\eta;
 \eeq   
  we obtain:
 \beq \label{LAM}
 \bar{\gamma}_\eta \,\,=\,\,\sqrt{2\,+\,\bas}\,-\,1; ~~~~\lambda_\eta\,\,=\, \,\h\,\frac{\bas}{3 \,+\,2\,\bas\,-\,2\sqrt{2\,+\,\bas}};
 \eeq
 Re-writing \eq{LAM}  for the saturation momentum  as a function of $Y$ we obtain
 
 \beq \label{LAMY}
 \ln\Lb \frac{Q^2_s\Lb Y\Rb}{Q^2_s\Lb Y=0\Rb} \Rb\,\,=\,\,\lambda \,Y
 ~~~\mbox{with}\,\,\lambda\,\,=\,\,\frac{\lambda_\eta}{1\,+\,\lambda_\eta}  ~~\mbox{and} ~~~\bar{\gamma} \,=\,\bar{\gamma}_\eta\Lb 1\,+\,\lambda_\eta\Rb.
 \eeq
\begin{figure}
\centering 
   \includegraphics[width=9cm]{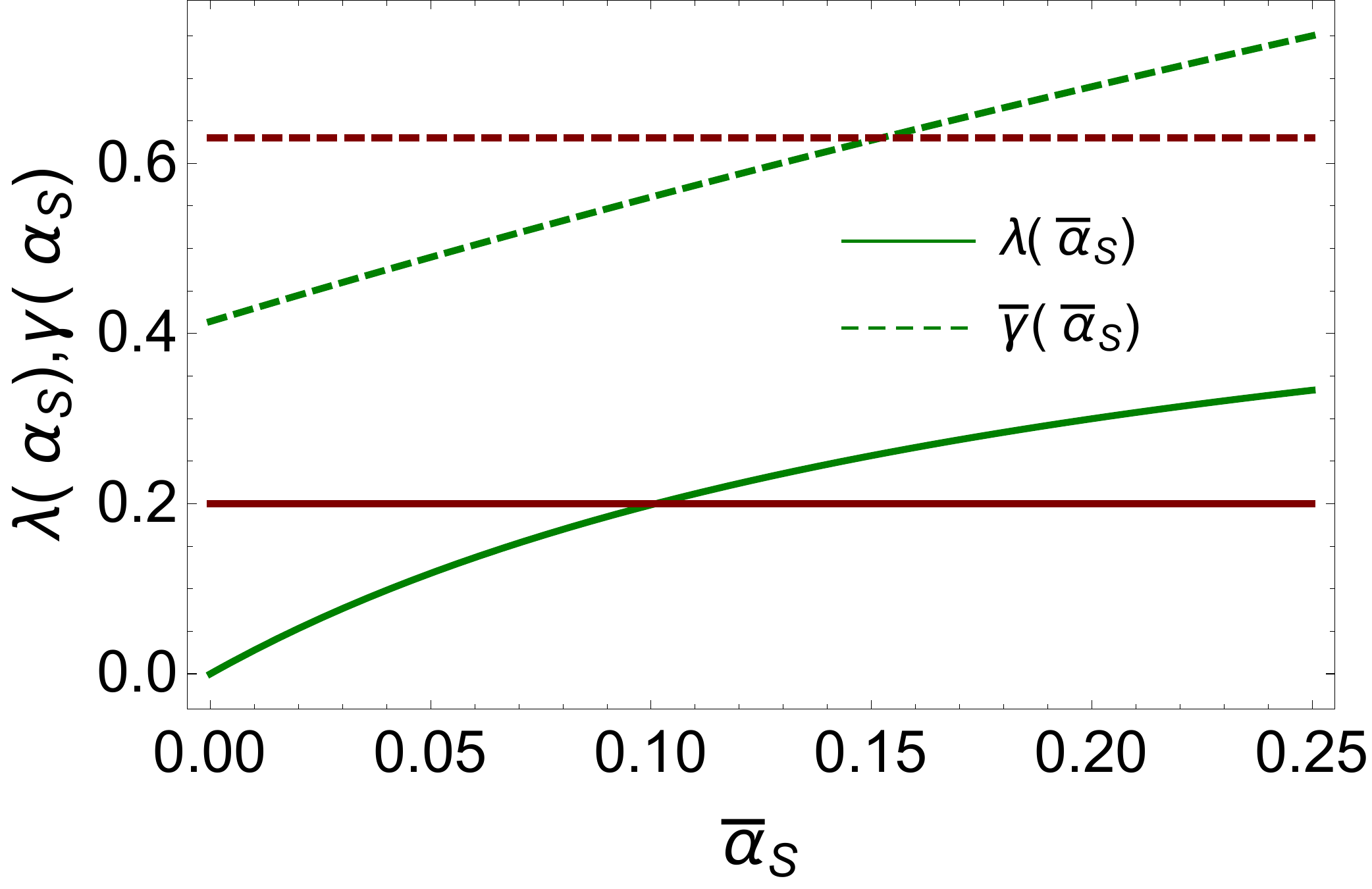}    
\caption{ $\lambda$ and $\bar{\gamma}$ versus $\bas$. In red we put $\bar{\gamma} = 0.63$ which stems from LO estimates and  $\lambda = 0.2$ which is a typical value that is need to describe DIS data (see  for example Ref.~\cite{SATMOD17}).}    
\label{lam}
\end{figure}


  \subsection{NLO BFKL kernel in the saturation domain}
  \begin{boldmath}
  \subsubsection{The kernel in $\gamma$-representation}
  \end{boldmath}
  
  The non-linear corrections are essential in the region of small values of $x$ (large $Y$) where $ \tau \,\,=\,\,r^2 Q^2_s \,\,>\,\,1$. In Ref.~\cite{LETU} it is shown that in the BFKL equation we have two types of logarithms, which are generated by the following kernels
\bea \label{SIMKER}
\chi\Lb \gamma\Rb\,\,=\,\, \left\{\begin{array}{l}\,\,\,\frac{1}{\gamma}\,\,\,\,\,\,\,\,\,\,\mbox{for}\,\,\,\tau\,=\,r Q_s\,>\,1\,\,\,\,\,\,\mbox{summing} \Lb \ln\Lb r Q_s\Rb\Rb^n;\\ \\
\,\,\,\frac{1}{1 \,-\,\gamma}\,\,\,\,\,\mbox{for}\,\,\,\tau\,=\,r Q_s\,<\,1\,\,\,\,\,\mbox{summing}
\Lb \ln\Lb1/(r\,\Lambda_{\rm QCD})\Rb\Rb^n;\\  \end{array}
\right.
\eea
In Ref.~\cite{CLMS}  we suggest to use \eq{SIMKER} instead of the full expression of \eq{CHI}.

In the previous section we specified how we changed the kernel in the
 perturbative QCD region, taking into account the NLO corrections.  In the region of large $\tau$   
 we need to find the anomalous dimension  at $\gamma\,\to\,0$, as it follows from \eq{SIMKER}.
Hence 
  we re-write \eq{KMRSOM}  in the vicinity of $\gamma \to 0$,  where it has the form:
   \beq \label{OMGA01}
 \omega\,\,=\,\,\bas \Bigg\{ \frac{1}{\gamma} \,\,-\,\,\omega\Bigg\}
 \eeq  
  Resolving \eq{OMGA01} with respect of $\gamma$ we obtain
  
  \beq \label{KERGA1}
  \gamma\,\,=\,\,\frac{\bas}{1 + \bas }\, \frac{1}{\omega}
  \eeq
  which differs from the behaviour of the LO kernel, only by  replacing 
 $\bas \,\to\,\bas/\Lb 1 \,+\,\bas\Rb$. Therefore, we can discuss the
 non-linear equation in the LO,   substituting $\bas/\Lb 1 \,+\,\bas\Rb$
 in place of $\bas$, at the final stage.  
    
  \subsubsection{The non-linear equation}
   In the saturation region 
 where $\tau\,\,>\,\,1$,  the  logarithms
   originate from the decay of a large size dipole, into one small
 size dipole  and one large size dipole~\cite{LETU}.  However, the size of the
 small dipole is still larger than $1/Q_s$. This observation can be
 translated in the following form of the kernel in the LO
\bea \label{K2}
\frac{\bas}{2 \pi}\int \, \displaystyle{K\Lb \vec{x}_{01};\vec{x}_{02},\vec{x}_{12}\Rb}\,d^2 x_{02} \,&\rightarrow&
\,\frac{\bas}{2}\, \int^{x^2_{01}}_{1/Q^2_s(Y,b)} \frac{ d x^2_{02}}{x_{02}^2}\,\,+\,\,
\frac{\bas}{2}\, \int^{x^2_{01}}_{1/Q^2_s(Y, b)} \frac{ d |\vec{x}_{01}  -
 \vec{x}_{02}|^2}{|\vec{x}_{01}  - \vec{x}_{02}|^2}\,\,\nn\\
 &=&\,\,\frac{\bas}{2}\, \int^{\xi}_{-\xi_s} d \xi_{02} \,\,+\,\,\frac{\bas}{2}\, \int^{\xi}_{-\xi_s} d \xi_{12}\eea
 where $\xi_{ik} \,=\,\ln\Lb x^2_{ik} Q_s^2(Y=Y_0)\Rb$
 and $\xi_s\,=\,\ln\Lb Q^2_s(Y)/ Q_s^2(Y=Y_0)\Rb$.

Inside the saturation region the BK equation of the LO  takes the form
\beq \label{BK2}
\frac{\partial^2 \widehat{N}\Lb Y, \x; \vec{b}\Rb}
{ \partial Y\,\partial \xi}\,\,=\,\, \bas \,\left\{ \Lb 1 
\,\,-\,\frac{\partial \widehat{N}\Lb Y, \xi; \vec{b}
 \Rb}{\partial  \xi}\Rb \, \widehat{N}\Lb Y, \xi;
 \vec{b}\Rb\right\}
\eeq
where 
 $\widehat{N}\Lb Y, \xi; \vec{b}\Rb\,\,=\,\,\int^{\xi} d \xi'\,N\Lb Y,
 \xi'; \vec{b}\Rb$ .
    
  For the NLO kernel of \eq{KERGA1},  \,  \eq{BK2} takes the form:
  \beq \label{BK3}
\frac{\partial^2 \widehat{N}\Lb Y, \xi; \vec{b}\Rb}
{ \partial Y\,\partial \xi}\,\,=\,\, \frac{\bas}{1\,+\,\bas} \,\left\{ \Lb 1 
\,\,-\,\frac{\partial \widehat{N}\Lb Y, \xi; \vec{b}
 \Rb}{\partial  \xi}\Rb \, \widehat{N}\Lb Y, \xi;
 \vec{b}\Rb\right\}
\eeq  
  
\subsubsection{The solution}

For solving this equation we introduce function $\Omega\Lb Y;
 \xi, \vec{b}\Rb$~\cite{LETU}
\beq \label{SOL1}
N\Lb Y, \xi \Rb\,\,=\,\,1\,\,-\,\,\exp\Lb - \Omega\Lb Y, \xi\Rb\Rb
\eeq
Substituting \eq{SOL1} into \eq{BK3} we  reduce it to the form

\begin{subequations}
\bea \label{SOL2}
&&\frac{ \partial \Omega\Lb Y, \xi\Rb}{ \partial Y} \,\,=\,\,\frac{\bas }{1 + \bas}\widetilde{N}\Lb Y, \xi\Rb;~~~\frac{ \partial^2 \Omega\Lb Y, \xi\Rb}{ \partial Y\,\partial \xi}  \,\,\,=\,\,\frac{\bas }{1 + \bas}\Bigg( 1 -\,\exp\Lb - \Omega\Lb Y, \xi\Rb\Rb\Bigg);\label{SOL02}\\
&&~\frac{ \partial^2 \Omega\Lb \xi_s; \zeta\Rb}{ \partial \xi_s\,\partial \xi}  \,\,\,=\,\,\frac{\bas}{\lambda(\bas)\,\Lb 1\,\,+\,\,\bas\Rb}\Bigg( 1 -\,\exp\Lb - \Omega\Lb \xi_s; \zeta\Rb\Rb\Bigg)
\,\,\equiv\,\,\sigma\Bigg( 1 -\,\exp\Lb - \Omega\Lb \xi_s; \zeta\Rb\Rb\Bigg)\label{SOL2}\eea
\end{subequations}
where $\lambda$  is given by   \eq{LAMY}. The variable $\xi_s$ is
 defined as 
\beq \label{XIS}
\xi_s\,\,=\,\,\ln\Lb Q^2_s\Lb Y\Rb/Q^2_s\Lb Y=0; \vec{b},\vec{R}\Rb\Rb\,\,=\,\,\lambda \,Y
\eeq
The use of this variable indicates the main idea of our approach in the region of $\tau \,>\,1$: we
 wish to match the solution of the non-linear \eq{SOL2} with the solution  of linear equation  (see \eq{DLALONL20}) in the kinematic region where it   has the form~\cite{MUT}: 
 \beq \label{Z} 
 N \,\,=\,\,N_0 \,\exp\Lb \bar{\gamma} \,z\Rb
~~~~\mbox{with} ~~~~~z\,\,=\,\,\xi_s\,\,+\,\,\xi
\eeq
and $\bar{\gamma}$ is determined by \eq{LAMY}. \eq{Z} leads to the initial and boundary conditions:

\beq \label{IC}
N\Lb r^2, Y\Rb\,\,=\,\,N\Lb \tau=1\Rb\,\,=\,\,N_0;~~~~~~~
\frac{d \ln \Lb N\Lb r^2, Y\Rb\Rb}{d \xi}|_{\tau=1} \,\,=\,\,\bar{\gamma};
\eeq
\eq{SOL2} has a traveling wave solution (see formula {\bf 3.4.1.1} of
  Ref.~\cite{MATH}). For \eq{SOL2} in the canonical form:
\beq \label{SOL3}
\frac{ \partial^2 \Omega\Lb \xi_s; \tilde{ \zeta}\Rb}{ \partial t^2_+} \,\,-\,\,\frac{ \partial^2 \Omega\Lb \xi_s; \tilde{\zeta}\Rb}{ \partial t^2_-} \,\,\,=\,\,\sigma\,\,\Bigg( 1 -\,\exp\Lb - \Omega\Lb \xi_s; \tilde{\zeta} \Rb\Rb\Bigg),
\eeq
with $t_{\pm} = \xi_s \pm  \xi $, the solution takes the form:
\beq \label{SOL4}
\int^{\Omega}_{\Omega_0}\frac{d \Omega'}{\sqrt{ C_1 + \frac{2}{(\mu^2 - \kappa^2)} \sigma \Lb \Omega' + \exp\Lb-\Omega'\Rb\Rb}}\,\,=\,\,\mu t_+ + \kappa t_-  + C_2
\eeq
where all constants have to be determined  from the initial and boundary 
conditions of \eq{IC}. First we see that $C_2=0 $ and $\kappa =0$.
 From the condition $\Omega'_z/\Omega \,\,=\,\,\bar{\gamma}$ at $t_+ =
 0$ we can find $C_1$. Indeed, differentiating \eq{SOL4} with respect to 
$t_+$ one
 can see that at $t_+= 0$ we have:
\beq \label{SOLSET}
\frac{d \Omega}{d t_+}|_{t_+ = 0}\,\frac{1}{\sqrt{ C_1 \,\,+\,\,\frac{2 \,\sigma}{\mu^2}\Lb 1\,\,+\,\,\h\,\Omega^2_0\Rb}}\,\,=\,\,\mu
\eeq

From     \eq{SOLSET} one can see that choosing 

\beq \label{SOLSET1}
C_1\,\,=\,\,- \,\frac{2 \sigma}{\mu^2} \,\,+\,\,\Omega^2_0 \Lb 1\,\,-\,\,\frac{ \sigma}{\mu^2}\Rb;~~~~~\mbox{and}\,~~~\mu \,\,=\,\,\bar{\gamma} 
\eeq

we satisfy the initial condition $\frac{d \ln\Lb \Omega\Rb}{d z}|_{t_+ = 0}
 =\,\bar{\gamma}$ of \eq{LAMY}.

Finally, the solution of \eq{SOL4} can be re-written in the following
 form for $\Omega_0 \ll 1$:
\beq \label{SOL5}
\int^{\Omega}_{\Omega_0}\frac{d \Omega'}{\sqrt{ \Omega^2_0\,\Lb1\,\,-\,\,\frac{\sigma}{\bar{\gamma}^2}\Rb\,\,+\,\,\frac{2\,\sigma}{\bar{\gamma}^2}\Lb -1\,\,+\,\,\Omega'\,\,+\,\,\,e^{ - \,\Omega'}\Rb
}}\,\,=\,\, \bar{\gamma}\,z \eeq

One can see that \eq{SOL5} gives the geometric scaling solution~\cite{LETU,BALE,IIML,SGBK} which depends only  on the variable $z$.
For $\Omega \to \Omega_0$ and if $\Omega_0 \ll 1$, \eq{SOL5} can be 
 solved explicitly giving
\beq \label{SOL7}
\Omega\,\,=\,\,\Omega _0	\Bigg\{ \cosh \left(\sqrt{\sigma } \Lb\xi_s + \xi \Rb \right)\,\,+\,\,\frac{\bar{\gamma}}{\sqrt{\sigma}}\, \sinh \left(\sqrt{\sigma}\Lb  \xi_s + \xi\Rb\right)\Bigg\}\eeq
 \eq{SOL7} gives the solution which depends only on one variable
 $z\,\,=\,\,\xi_s + \xi$,
and satisfies the initial conditions of \eq{IC}.

At large $z$ we obtain the solution~\cite{LETU}:
\beq \label{SOL8}
\Omega\Lb z\Rb\,\,=\,\,\frac{\sigma}{2}\,z^2\,\,+\,\,{\rm Const}
\eeq
or in terms of the amplitude
\beq \label{SOL9}
N\Lb z\Rb\,\,=\,\,1\,\,-\,\,{\rm Const}\,e^{-\,\frac{\sigma}{2}\,z^2}
\eeq
  We wish  to stress that  \eq{SOL9} reproduces the 
 asymptotic solution to the BK equation in the NLO, which has been derived in
 Refs.~\cite{CLMP,XCWZ}, for fixed $\bas$.

It should be noted that both solutions of \eq{SOL7} and \eq{SOL8}
 can be derived directly from \eq{SOL3} assuming $1\,-\,\exp\Lb -
 \Omega\Rb\,\,\to\,\Omega$ and $1\,-\,\exp\Lb - \Omega\Rb\,\,\to\,1$ for
 small  $z$ and large $z$, respectively.

\section{Our approach}
 
\subsection{Generalties}
\begin{figure}[ht]
    \centering
  \leavevmode
      \includegraphics[width=8cm]{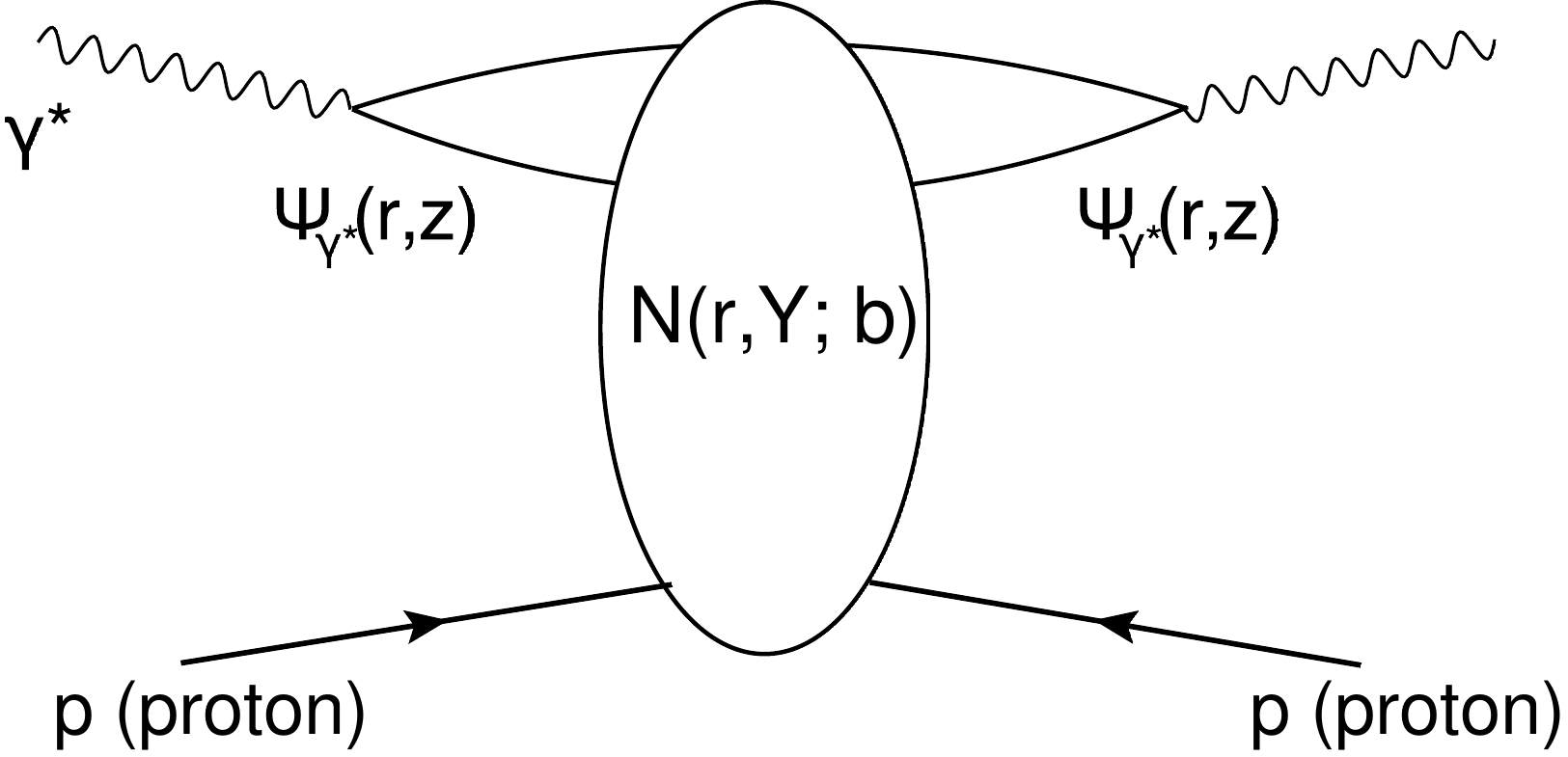}  
      \caption{The graphic representation of \protect\eq{FORMULA} for the scattering amplitude. $Y = \ln\Lb 1/x_{Bj}\Rb$ and $r$ is the size of the interacting dipole.   $z$  denotes  the fraction of energy that is carried by one quark. $b$ denotes  the impact parameter of the scattering amplitude.}
\label{gen}
   \end{figure}

   The observables in deep inelastic scattering can be expressed through the following scattering amplitudes (see \fig{gen} and Ref.  ~\cite{KOLEB} for the review and references therein)

\beq\label{FORMULA}
N\Lb Q, Y; b\Rb \,\,=\,\,\int \frac{d^2 r}{4\,\pi} \int^1_0 d z \,|\Psi_{\gamma^*}\Lb Q, r, z\Rb|^2 \,N\Lb r, Y; b\Rb
\eeq
where $Y \,=\,\ln\Lb 1/x_{Bj}\Rb$ and $x_{Bj}$ is the Bjorken $x$. $z$ is the fraction of energy carried by quark.
$Q$ is the photon virtuality. $b$ denotes  the impact parameter of the scattering amplitude.

\eq{FORMULA} shows the main features of the interactions at high energies. They go in two stages. The first is the decay of virtual photon in quark-antiquark pair, described by  $|\Psi_{\gamma^*}\Lb Q, r, z\Rb|^2$ in \eq{FORMULA}.  For large $Q^2$ ( $Q^2 \,\geq\,Q^2_0$ with $Q^2_0 \approx 0.7 GeV^2$,  see Ref.  ~\cite{GLMTC}) the wave function is well known (see Ref. ~\cite{KOLEB} and references therein)
\begin{align}
  (\Psi^*\Psi)_{T}^{\gamma^*} &=
   \frac{2N_c}{\pi}\alpha_{\mathrm{em}}\sum_f e_f^2\left\{\left[z^2+(1-z)^2\right]\epsilon^2 K_1^2(\epsilon r) + m_f^2 K_0^2(\epsilon r)\right\},\label{WFDIST}   
  \\
  (\Psi^*\Psi)_{L}^{\gamma^*}&
  = \frac{8N_c}{\pi}\alpha_{\mathrm{em}} \sum_f e_f^2 Q^2 z^2(1-z)^2 K_0^2(\epsilon r),
\label{WFDISL}
\end{align}
where T(L) denotes the polarization of the photon and $f$ is the flavours of the quarks. $\epsilon^2\,\,=\,\,m^2_f\,\,+\,\,Q^2 z (1 - z)$. However, even for DIS the non-perturbative corrections become essential and we cannot use \eq{WFDIST} and \eq{WFDISL} for $Q^2 \leq Q^2_0$.

The second is the amplitude $N\Lb r, Y; b\Rb$
of the interaction of the dipole with the target. We have discussed this amplitude in the previous section and will specify the way how we will use the finding of  this section  for the practical application to DIS below.

 Using \eq{FORMULA}, \eq{WFDIST} and \eq{WFDISL}  we can write the main observables in DIS:
 \begin{subequations}
 \bea 
 \sigma_{T,L} &\,\, =\,\,& 2 \int d^2 b \,\,N_{T,L}\Lb Q,Y,;b\Rb; \label{SIGMA}\\
 F_2\Lb Q, Y\Rb &=& \frac{Q^2}{4\,\pi^2\,\alpha_{\rm e.m.}} \{ \sigma_T\,\,+\,\,\sigma_L\};\label{F2}\\
  F^{cc}_2\Lb Q, Y\Rb &=& \frac{Q^2}{4\,\pi^2\,\alpha_{\rm e.m.}} \{ \sigma^{cc}_T\,\,+\,\,\sigma^{cc}_L\};\label{F2CC}\\
  F_L\Lb Q, Y\Rb &=& \frac{Q^2}{4\,\pi^2\,\alpha_{\rm e.m.}} \,\sigma_L;\label{FL} 
 \eea
  \end{subequations} 
 $\sigma^{cc}$ in \eq{F2CC} are calculated, using \eq{SIGMA} with  $N_{T,L}\Lb Q,Y,;b\Rb$, which is calculated using \eq{WFDIST} and \eq{WFDISL} for $c$-quarks.
\subsection{The dipole scattering amplitudes}
  Based on the approach, briefly discussed in the previous section, we   can see three  different kinematic regions where we are going to use the scattering amplitude $ N\Lb Y,
 \xi'; \vec{b}\Rb$ in three different forms.

  \begin{enumerate}
  \item\quad  For $ Q^2_s\,r^2\,\,\,<\,\,1$ (perturbative QCD region) 
  we suggest to use the linear evolution equation of \eq{DLALONL20}, which has been discusssed above.
 Recalling 
that in the  derivation of this equation we use the DLA in which
 both  $\eta\,\,=\,\,Y\,-\,\xi' $   and $\xi'$ are considered to
 be large: $ \bas\, \eta\, \xi \,\gg\,\,1$.  For $\eta \,>\,0$
 ($ Y\,>\,\xi$) we can use the experimental data as the initial
 condition for \eq{DLALONL20}.    The value of the saturation momentum
 is given by \eq{LAMY}. Since its value turns out to be much
 less than $Q_{\rm max}$, which stems from the condition $\xi_{max}
 = \ln \Lb Q^2_{\rm max}/Q^2_0\Rb$\,\,=\,\,$\eta$, we can use the DGLAP evolution
 equation in the next-to-leading order for  $Q^2 \,>\,Q^2_{\rm max}$.

        \item\quad  For $ Q^2_s\,r^2\,\,\,\sim\,\,1$ (vicinity of the
 saturation scale) we use the scattering amplitude  in the form ~\cite{MUT}
        \beq \label{MF2}     
        N\Lb r,   \eta ; b \Rb \,\,=\,\,N_0\,\,\Lb  Q^2_s\Lb Y, b \Rb\,r^2\Rb^{\bar{\gamma}}
        \eeq
        with $\bar{\gamma}$ from \eq{LAMY}.
 \item\quad  For $ Q^2_s\,r^2\,\,\,\gg\,\,1$ (saturation region), we 
propose
 to use the solution to the non-linear equation of \eq{SOL5}.
 \end{enumerate} 
 
 For $\tau \,<\,1$ we have to solve the linear equation with some initial condition which we need to take from the experimental data.  This solution will give us the scattering amplitude in the vicinity of the saturation scale where the scattering amplitude can be estimated using \eq{MF2}. We use a different way to introduce the parameters from the experimental data: we expand \eq{MF2} to the region $\tau \,<\,1$ , replacing $\bar{\gamma}$ by following expression:
 
 \beq \label{GAEFF}
 \bar{\gamma}\,\,\longrightarrow\,\, \bar{\gamma}\,\,\,+\,\,\frac{\ln\Lb1/\tau\Rb}{\kappa\,\lambda\,Y};
 ~~~\mbox{with}\,\,\kappa \,\,=\,\,\frac{  \frac{d^2 \omega\Lb \bar{\gamma}_\eta \Rb}{d \bar{\gamma}_\eta ^2}}{\frac{d \omega\Lb \bar{\gamma}_\eta \Rb}{d \bar{\gamma}_\eta }}
 \eeq
 
 \eq{GAEFF} is derived in Ref. ~\cite{IIML} and the experience with the saturation models~\cite{SATMOD0,SATMOD1,SATMOD2,IIM,SATMOD3,SATMOD4,SATMOD5,SATMOD6,SATMOD7,SATMOD8, SATMOD9,SATMOD10,SATMOD11,SATMOD12,SATMOD13,SATMOD14,SATMOD15,SATMOD16,SATMOD17,CLP,CLMP} shows that it is described the experimental data for $x\,\leq\,0.1$ quite well. Hence, in our approach the values of all phenomenological parameters of  \eq{MF2} (see below)    should be determined from the experimental data.
 
 The mass of the $c$-quark (about  $m_c$ = 1.4\,GeV) is not small and we took this into account replacing $x$ in the $c  \bar{c}$ scattering by
 \beq \label{XCC}
 x^c\,\,=\,\,\Lb 1\,\,+\,\,\frac{4\,m^2_c}{Q^2}\Rb x
 \eeq
 
 For $ \tau\,\,=\,\,Q^2_s\,r^2\,\,\,\gg\,\,1 $ we need to use  the solution of \eq{SOL5}  to the non-linear evolution equation. However, it has been found in Ref. ~\cite{LEPP} that the following formula:
 \beq \label{APRAM}
 N\Lb z\Rb\,\,=\,\,a\,\Big( 1\,\,-\,\,\exp\Lb - \Omega\Lb z \Rb\Rb\Big) \,\,+\,\,\Lb 1\,-\,a\Rb\frac{\Omega\Lb z \Rb}{1\,\,+\,\,\Omega\Lb z \Rb} 
\eeq 
 with $\Omega\Lb z \Rb$ from \eq{SOL7}  and with $a$  = 0.65 describes the exact solution within accuracy less that 2.5 \%. Therefore, we use \eq{APRAM} in our attempts to describe the HERA data.
 
\subsection{Phenomenological input}
 As has been mentioned above, we need to set the initial conditions at $Y=0$  for the linear evolution equation for $\tau\,\leq\,1$. In this section we wish to clarify how we introduce the phenomenological parameters to describe this conditions. The first one we have considered: $N_0$ which determines the scattering amplitude at $\tau\,=\,  1$. Two other parameters  describe the saturation momentum at $Y=0$: the value of  $Q_s$  at $b$ = 0 and the behaviour as a function of $b$. collecting everything that we have discussed about $Q_s$ we  use two expression for the saturation momentum:
  \begin{subequations}
 \bea  
 1.~~Q^{(1)2}_s\Lb Y, b\Rb\,\,&=&\,\,Q^{(1) 2}_s\Lb Y=0,b=0\Rb\,\,e^{ - m\,b}\,\,e^{\lambda\,Y} \,\,=\,\,Q^{2}_0\,e^{ - m\,b}\,e^{\lambda\,Y} ;\label{QS1}\\
 2.~~Q^{(2)2}_s\Lb Y, b\Rb\,\,&=&\,\,Q^{(2) 2}_s\Lb Y=0,b=0\Rb\,\,e^{ - \frac{3}{4}\,{\cal Z}}\,\,e^{\lambda\,Y} \,\,=\,\,Q^{2}_0\,\,e^{\lambda\,Y} \,e^{ - \frac{3}{4}\,{\cal Z}} ;\label{QS2}
 \eea
 \end{subequations}
 where ${\cal Z}$ and $\lambda$ are  defined in \eq{QSB1} and \eq{LAMY}, respectively.  The values of these parameters have to be found from  fitting of the experimental data. The experience with such fitting gives $Q^{(1) 2}_0\,=\,Q^{(2) 2}_0 = 0.15 \div 0.25\,GeV^2$. $m$ we can estimate, assuming that $< b^2>\, =\,6/m^2 \,=\, R^2$ where $R$ is the electro-magnetic radius of the proton.
 Hence we expect $m \,\approx 0.55\,GeV$ but we have to remember that $m$ characterizes  the distributions of the gluons in the proton which can be quite different from the quarks\footnote{\label{ftnt}Actually, the above estimates should be made for the scattering amplitude $N = N_0(r^2 Q^2_s)^{\bar{\gamma}}$,  which leads to $m \approx\,\,0.34\,GeV$: with this estimates we need to compare  the value of $m$ from the Table I}. ${ \alpha'}^2_{\rm eff} $ in ${\cal Z}$ has been evaluated in Ref.~\cite{LEPION} with $ { \alpha'}^2_{\rm eff} \approx 0.1\,GeV^{-2}$ for $\bas = 0.2$.
 
 Finally, we introduce three phenomenological parameters  from the initial conditions whose values have to be found from the fit of the experimental data. The masses of quarks, that determine the wave function of the virtual photon,  determine the infra-red behaviour of the wave function. We take two sets of them:  the current masses and the masses of light quarks are equal to 140\,MeV which is the typical infra-red cutoff in our approach.
 
\section{Results of the fits}
 Using the approach, that has been discussed in the previous section, we attempt to describe the most accurate data for the deep inelastic structure function $F_2$ ~\cite{HERA1}. The implicit parameters of our approach are the restrictions of the kinematic region of the experimental data that we include in the fit. We chose: $ 0.85\,GeV^2 \leq Q^2\leq 27\,GeV^2$ and  $x \,\leq \,0.01$. The lower limit of $Q^2$ stems from non-perturbative correction to the wave  function of the virtual photon, while the upper limit originates  from two  restrictions:  $x\,\leq\,0.01$ 
  and contribution of the additional term in $\bar{\gamma}$ of \eq{GAEFF} is small.  The choice of the largest $x$ is dictated by the  needs to have low $x$ for legitimate use of our  theoretical formulae and the practical wish to use in the fit as more data as possible.


\begin{table}[ht]
{\footnotesize
\begin{tabular}{||l|l|l|l|l|l|l|l|l|l|l||}
\hline
\hline
 &\multicolumn{5}{c|}{Dipole amplitude} & \multicolumn{4}{c|}{Wave function}& $\chi^2/d.o.f.$ \\
 \hline
 Set & $\bas$ & $N_0$ & $Q^2_0 (GeV^2)$ &m(GeV)&$\alpha_{eff}(GeV^{-2})$
 & $m_u $(MeV) &  $m_d $(MeV) &  $m_s $(MeV)&  $m_c $(GeV)  &0.85 $\leq Q^2\leq$ 27 $GeV^2$  \\
\hline
1 &  0.091 & 0.236& 0.998&   0.612 &n/a & 140 &140& 140&1.4& 124.9/133 = 0.93 \\
\hline
2 & 0.20 & 0.25 & 1.00&0.551&n/a & 140 & 140& 140 & 1.4 & 61.99/66  =0.93\\
\hline
3 & 0.096 & 0.448 & 0.921&0.840&n/a & 2.3 & 4.8& 95 & 1.4 & 117.2/133 = 0.88\\
\hline
4 & 0.20 & 0.343 & 0.999&1.300& n/a& 2.3 & 4.8& 95 & 1.4 & 91.74/66= 1.39\\
\hline
5& 0.038&0.599& 1.284 &n/a&0.216 &140 &140&140&1.4 & 175/133 = 1.31\\
\hline
6& 0.043&0.565&1.429 &n/a&0.149 &2.3 &4.8&95&1.4 &143.7/133=1.08\\\hline\hline
\end{tabular}}
\caption{Parameters of the model.    $\bas $, $N_0$,  $m(\alpha_{eff})$ and  $Q^2_0$ are fitted parameters. Two sets of quark masses are chosen: the current masses (sets 3 and 4) and the masses of light quarks  are equal to  $140 \,MeV$ (sets 1 and 2), which is the typical infra-red cutoff in our approach.  For the sets 2 and 4 the value of $\bas = 0.2$ is fixed and only three parameters were chosen from the fit.}
\label{t1}
\end{table}

     \begin{figure}[ht]
    \centering
  \leavevmode
      \includegraphics[width=18cm]{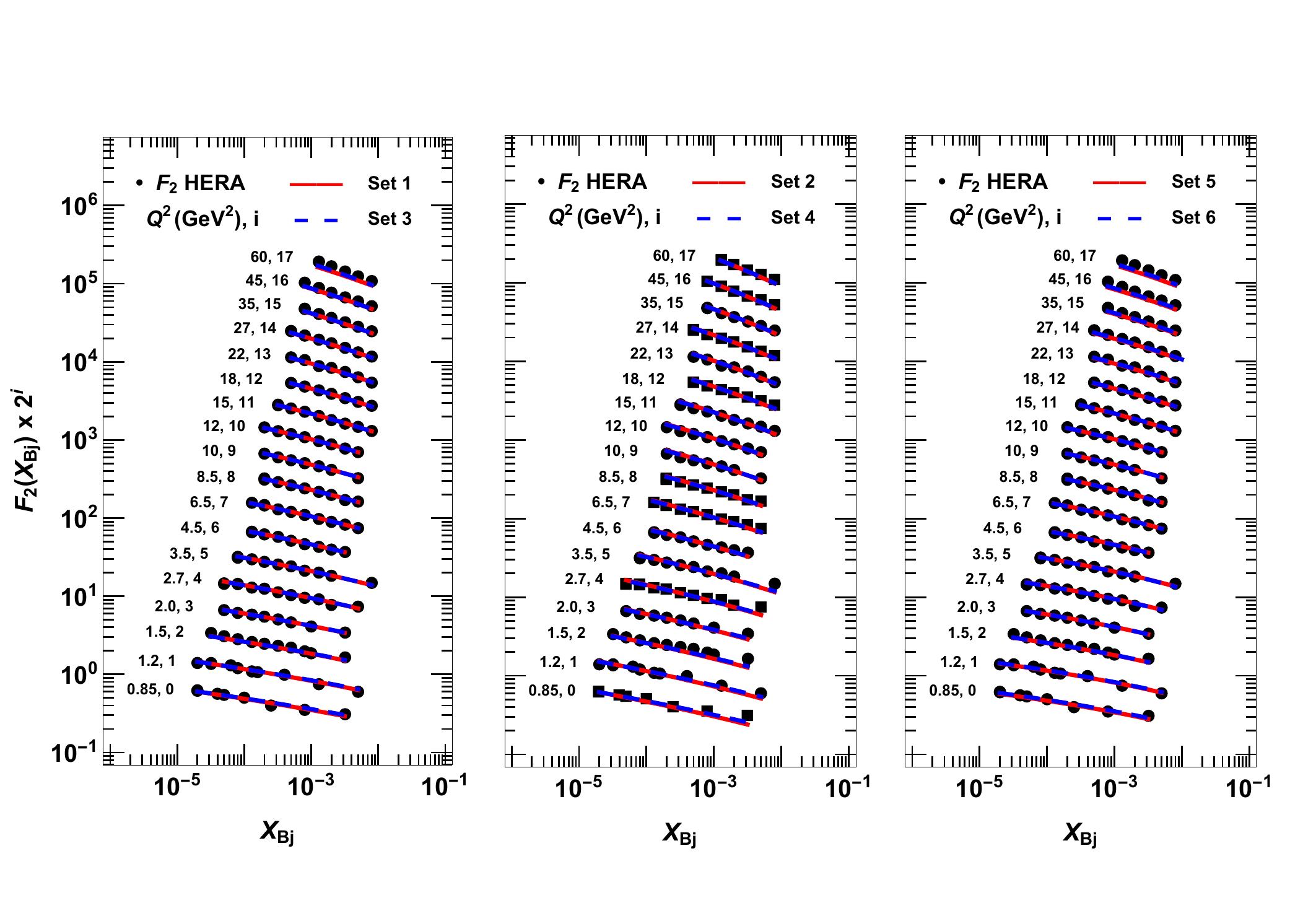}  
      \caption{Comparison with the experimental data of Refs.~\cite{HERA1,HERA2} of the sets shown in table I. The experimental data as well as the theoretical estimates are multiplied by factor $2^i$ and the values of $i$ are specified in the figures. In the panel with sets 2 and 4 we use  black squares to mark the experimental data that have been used in the fit.}
\label{f2hq}
   \end{figure}



\begin{figure}[h]
    \centering
  \leavevmode
      \includegraphics[width=12cm]{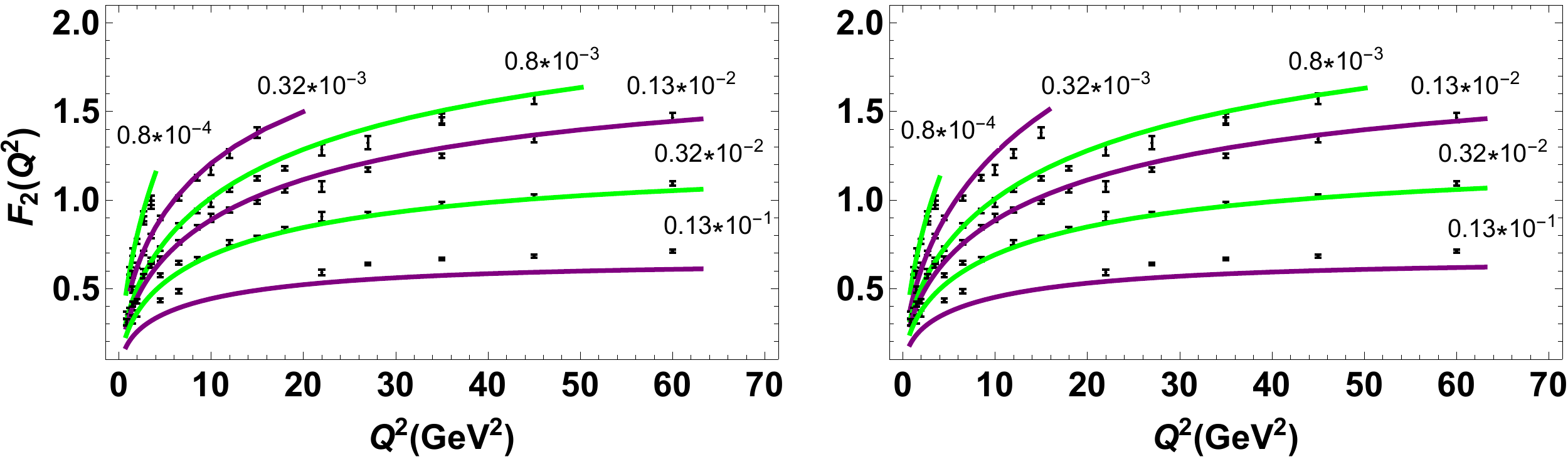}  
      \caption{$F_2$ versus $Q^2$ at fixed $x$ for the set 2 (left panel) and set 4 (right panel).}
\label{f2vsQ}
   \end{figure}

In Table 1 we show the  values of  parameters  from  our fits. It should be noted that we did not fit the mass of the quarks but use two sets of them: the masses of light quarks are equal to $m=\,140\,MeV$, which we consider as the typical infra-red cutoff in  our approach (sets 1 and 2); and the current masses (sets 3 and 4).
  We used the data for the inclusive $F_2$ in the kinematic region:  $ 0.85\,GeV^2 \leq Q^2\leq 27\,GeV^2$ and  $x \,\leq \,0.01$, for finding the parameters while   $F^{cc}_2$ as well as $F_2$ outside of this kinematic region (for  small values of $Q^2\,\leq\,0.85\,GeV^2$ or $Q^2\,\geq\,27 \,GeV^2$), are compared with the experimental data, to   demonstrate our ability to describe the experimental data. The quality of the fit one can see from \fig{f2hq}.  In \fig{f2vsQ} we show the data on $F_2$ as function of $Q^2$ at different values of $x$. This figure illustrates that we can describe the data for $x\,\leq 0.013$. In addition, we show in \fig{f2hq}   $F_2$ for $27\,\leq\,Q^2 \,\leq 60\,GeV^2$. The quality of the fit is not very good due to the small experimental errors. Indeed, $\chi^2/d.o.f.$  for the range of $Q^2$ from $0.85 \,GeV^2$ to $60\,GeV^2$ is equal to 2.16 (set 1), 1.93 (set 3) and 3.19 (set 5).  The reason for this  stems from the approximate character of \eq{GAEFF}.  The contributions of the nonlinear corrections are negligible for such large photon virtualities and, therefore,  our approach coincides with the one of Ref.\cite{DIMST} (see  also Ref.\cite{IMMST}), which describes the data  very well. We believe, that we demonstrated that our approach can describe the experimental data at rather small values of $Q^2$ where the non-linear corrections give an essential contribution. For large $Q^2$ our approach coincides  with the one of Ref.\cite{DIMST} which takes into account the linear evolution in much better way than we\cite{IMMST}. 
   
\begin{figure}[h]
    \centering
  \leavevmode
      \includegraphics[width=18cm]{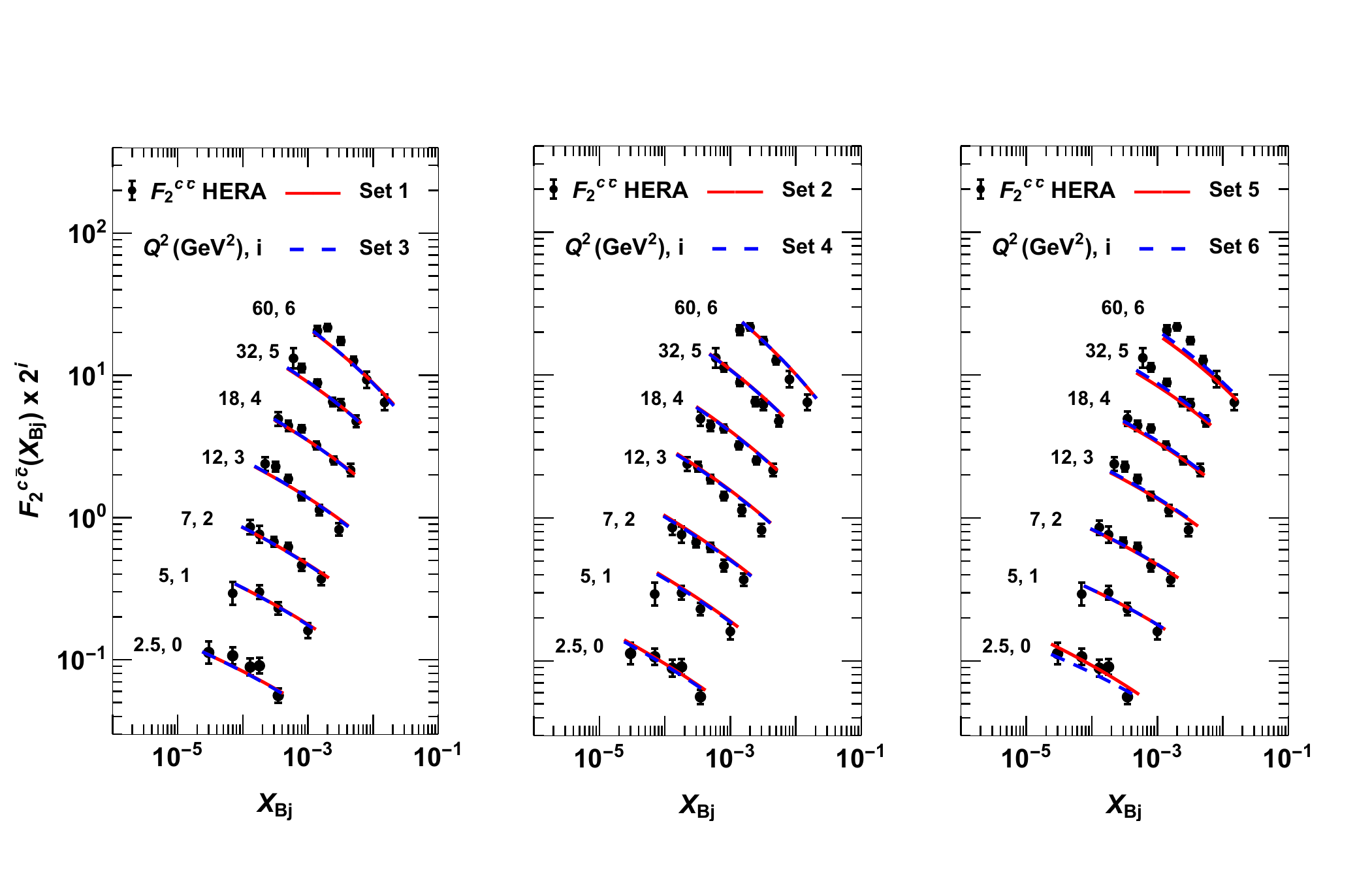}  
      \caption{Comparison our fits  with the experimental data  on $F^{cc}_2$  (see  Refs. ~\cite{HERA1,HERA2}) of the sets  shown in the table I. The experimental data as well as the theoretical estimates are multiplied by factor $2^i$ and the values of $i$ are specified in the figures. }
\label{f2cchq}
   \end{figure}

\begin{figure}[ht]
    \centering
  \leavevmode
      \includegraphics[width=12cm]{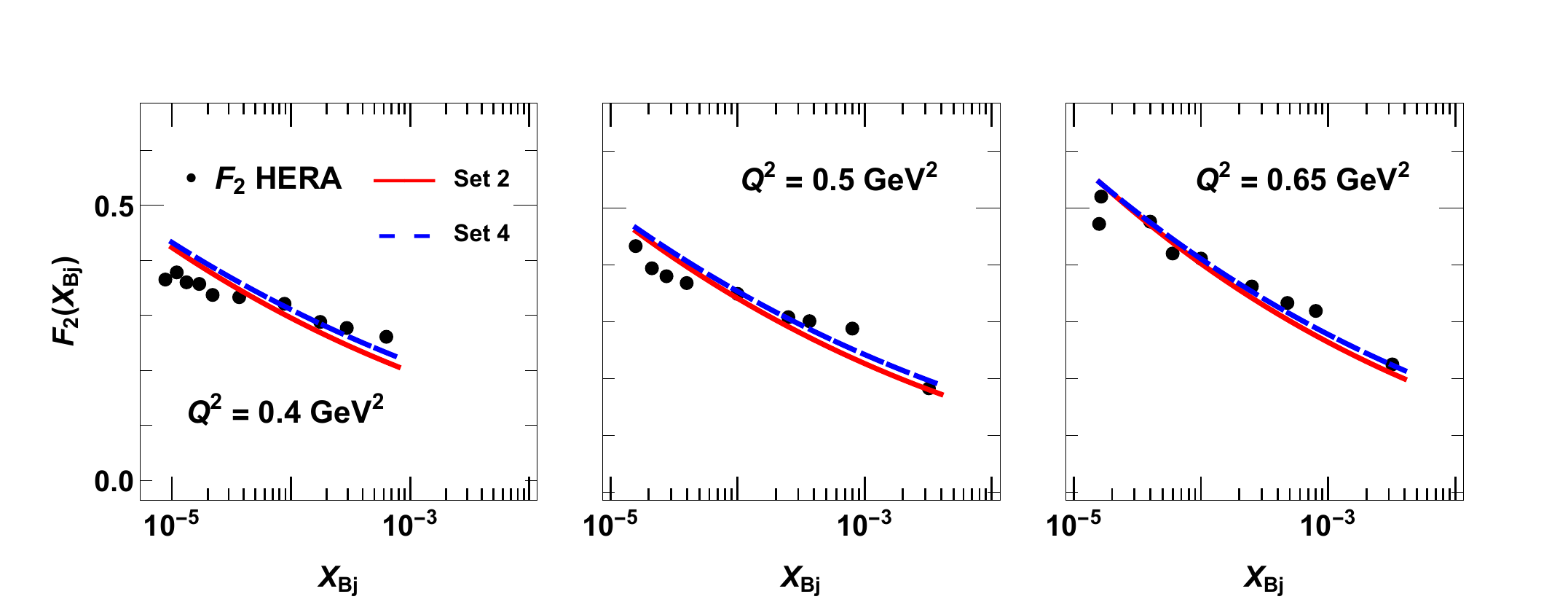}  
      \caption{$F_2$ versus $x$ at low $Q^2$. Red line for set 2 while the blue one for set 4.}
\label{f2Q}
   \end{figure}

  \fig{f2cchq}  demonstrates how our fits describe the experimental data on $F^{cc}_2$ while \fig{f2Q} shows that our fit is able to reproduce the data at low values of  $Q^2$. One can see that  the agreement with the experimental data is good even at low $Q^2$. It should be stressed that two sets of the light quark mass give the same description illustrating a possibility to use the wave function of the virtual photon in perturbative QCD at rather low values of $Q^2$.
  
\begin{figure}[ht]
    \centering
  \leavevmode
      \includegraphics[width=14cm]{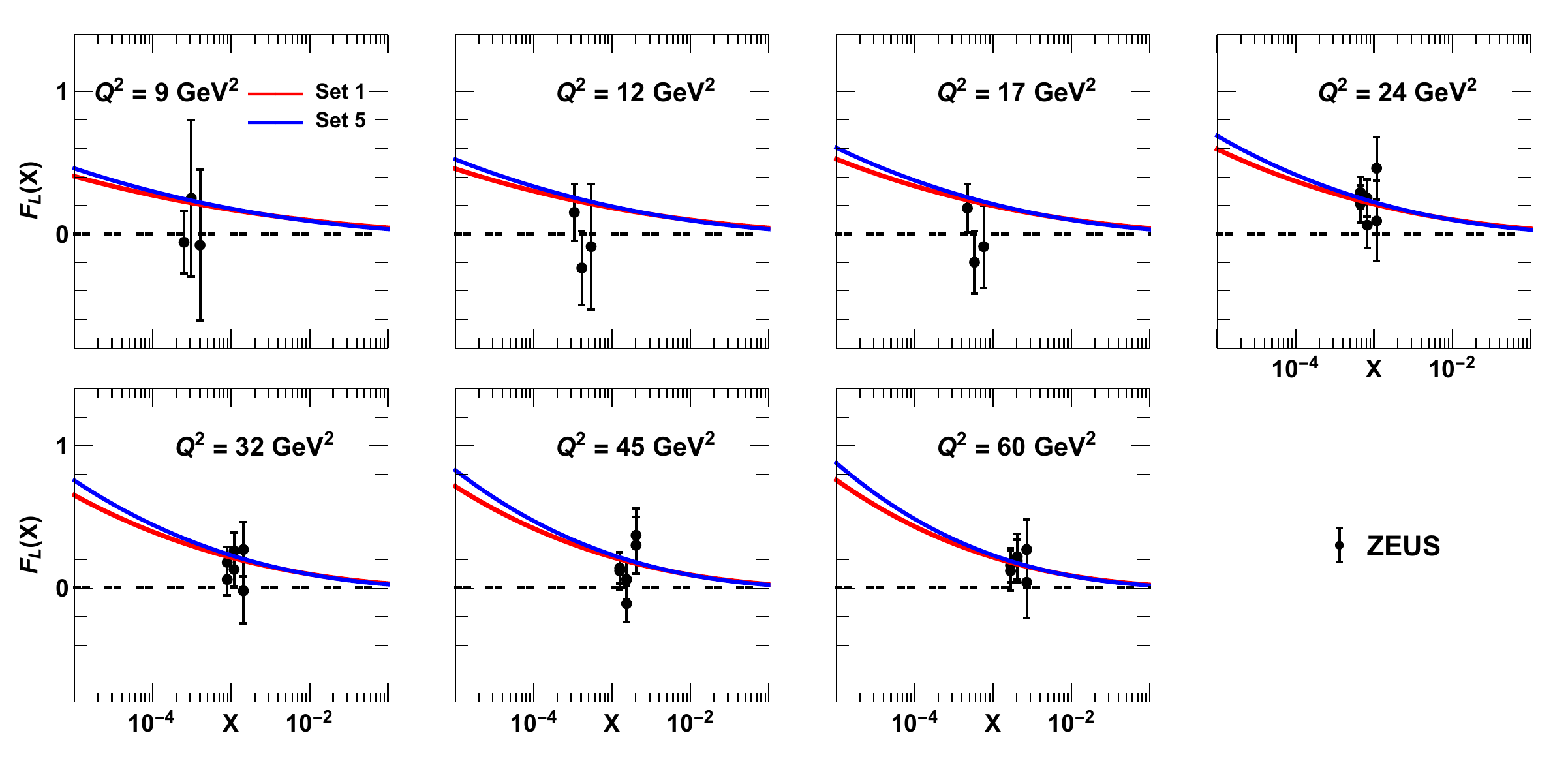}  
      \caption{ Comparison our fits  with the experimental data  on $F_L$  (see  Ref. ~\cite{HERAFL2}) of the sets  1 and 5. All other sets lead to the curves which are very close to these two. It is shown the data, extracted from the unconstrained fit (see Ref.\cite{HERAFL2} for explanation). Note that the data for constrained fit give $F_L\,\geq \,0$. All points with negative values of $F_L$ in the constrained fits coincide with 0.}
\label{fl}
   \end{figure}

  In addition to $F_2$ and $F^{CC}_2$ we compare  our approach with the experimental data on $F_L$\cite{HERAFL1,HERAFL2}. In spite of the fact that the data  have sufficiently large errors one can see from \fig{fl} that we are able to describe the data quite well. 
  
  Sets 2 and 4 need special comments. In these sets we choose $\bas = 0.2$ to illustrate the ability of our approach to overcome the difficulties of next-to-leading approaches in describing the experimental data: the small value of needed $\bas$.     One can see, that we obtain a good  $\chi^2/d.o.f.$ but for the specific set of data in the wide kinematic region $ 0.85 \leq\,Q^2\,\leq\,60\,GeV^2$ but choosing  a restricted selection  of data. Note that these data are marked by    the squares in the panel of   \fig{f2hq}   with sets 2 and 4.
  
 The general characteristics of our fits are: (1) the possibility to describe the data at rather  small values of $Q^2$, where the non-linear corrections turn out to be essential, (2) the large value of $Q^2_0  = 1\,GeV^2$;  (3) rather large value of $m$ in comparison with the estimates from the electro-magnetic radius of the proton (see above in footnote \ref{ftnt}); (4) the possibility to fit the experimental data,  taking into account the Gribov's diffusion  for impact parameter dependence of the saturation scale (see panel with sets 5 and 6 in \fig{f2hq}); and (5) an ability to describe the data at rather large values of $\bas$.

 Unfortunately, in spite of large numbers of the papers in which the experimental data have been compared with the saturation models
~\cite{SATMOD0,SATMOD1,SATMOD2,IIM,SATMOD3,SATMOD4,SATMOD5,SATMOD6,SATMOD7,SATMOD8, SATMOD9,SATMOD10,SATMOD11,SATMOD12,SATMOD13,SATMOD14,SATMOD15,SATMOD16,SATMOD17,CLP,CLMP} we can confront our approach only with Refs.~\cite{CLMP,CLP} since in all other papers the assumptions were made  that contradict the theoretical informations. For example, in the most models $Q^2_s$ is proportional to $\exp\Lb - b^2/B\Rb$ while this behaviour disagrees   with the Froissart theorem~\cite{FROI}; or/and the behaviour for $r^2 Q^2_s \,\,\gg\,\,1$ contradicts the approach of Ref.~\cite{LETU}. Comparing our parameters with Ref.~\cite{CLMP} one can see that the typical value of $\bas  \approx 0.14-0.15$ in Ref.~\cite{CLMP} is less than in sets 2 and 4. The value of $Q^2_0$  is about in three times larger than in this paper.
 The value of $m $ in Ref.~\cite{CLMP} is the same as in the electro-magnetic form factor of  proton, while in this paper $m$ is  larger, than in the electro-magnetic form factor of  proton in agreement with the  new experimental  information on the behaviour of the two gluon form factors ~\cite{KHAR,MAZA}. Taking into account that $\chi^2/d.o.f.$  are better  in this paper than in Ref.~\cite{CLMP,CLP}, we believe that the theoretical approach of  section II, give  reliable description of the current experimental data on DIS.

  We wish to draw your attention to the sets 5 and 6, which take into account the different dependence of the saturation momentum on $b$ (see  \eq{QS2}). From \fig{f2hq} and \fig{f2cchq} one can see that we are able to describe the experimental data. Hence we see that the suggested dependence on rapidity  of the impact parameter distribution  is in accord with the experimental data.
        
    \section{Conclusions}

   In this paper we compare with the experimental data   the NLO approach  of Ref.~\cite{CLMS}. In  this  approach  we include 
 the re-summation procedure, suggested in
 Refs. ~\cite{SALAM,SALAM1,SALAM2,DIMST}, to fix the BFKL kernel in the NL0, but  we treat differently the non-linear corrections. The advantage of our treatment is that we reproduce the correct  asymptotic behaviour
 at large $\tau = r^2 Q^2_s$. 
 
  Fixing our phenomenological parameters from the fit to the experimental data on  $F_2$  DIS structure function in the region: $0. 85 \leq Q^2\leq 27 \,GeV^2$ and $x \leq 0.013$, we found that can describe these data with very good $\chi^2/d.o.f.$. It should be stressed, that for the first time it is demonstrated that the impact parameter dependence of the saturation scale, which is given by \eq{QS2},  is an agreement with the data.   We believe that this is an interesting result since \eq{QS2} follows  from the first attempt to take into account both the diffusion on $\ln(k_T)$, which stems from perturbative QCD, and the Gribov's diffusion in $b$  ~\cite{GOLEB}.
  
  Using the  phenomenological parameters, that we found from the fit of $F_2$, we demonstrated that  our approach is able to describe the experimental data both on $F^{cc}_2$ and on $F_2$ at low values of $Q^2 =0.4 - 0.7 \,GeV^2$.  We also reproduce the experimental data on $F_L$.

  It should be pointed out that our description in the NLO is better or of the same quality as the description by LO with more fitting parameters.  Two problems,  that we faced in describing the data in the NLO approach, have been partly healed: we are able to fit the data with not small  value of $\bas  = 0.2$ and the value of the saturation momentum at $Y=0$ is about   $Q^2_0 \, \approx\,1\,GeV^2$ instead of $3 \,GeV^2$ of Ref. ~\cite{CLMP}.  However, we are aware that $Q_0 = 1\,GeV$ is rather large  (see for example Ref.~\cite{DKLN}).
  
  Concluding we wish to mentioned that the model that we developed here, is the only one which includes (i)
  the re-summed NLO corrections; (2) the correct  theoretical behaviour at large $\tau$; (3) the large $b$ dependence of the saturation scale in accord with the Froissart theorem as well as the non-linear evolution.
  We believe that it will be useful for further discussion of the high energy scattering in QCD.

\section{Acknowledgements}
   We thank our colleagues at Tel Aviv university and UTFSM for
 encouraging discussions. Our special thanks go  to E. Gotsman  for all 
his remarks and suggestions on this paper, and  to Yuri Ivanov  for technical support of the USM HPC cluster. M.S. thanks  M. Arriagada and C. del Valle from Universidad de la Frontera,
and   S. Nauto and J. Vidal from Universidad de Playa Ancha, as well as M. Siddikov from UTFSM  for their valuable advices on programming.

This research was supported  by 
 ANID PIA/APOYO AFB180002 (Chile),  Fondecyt (Chile) grants  
 1180118 and 1191434,  Conicyt Becas (Chile)  and PIIC 009/2021, DPP, Universidad T\'ecnica Federico Santa Mar\'ia.

\end{document}